# Quantitative predictions of photo-emission dynamics in metal halide perovskites *via* machine learning

John M. Howard[1,2], Qiong Wang[3], Erica Lee[1,2],
Richa Lahoti[2,5,6], Tao Gong[6], Meghna Srivastava[6], Antonio Abate[3,4], Marina S. Leite[6]

[1]Department of Materials Science and Engineering, University of Maryland, College Park, MD 20742, USA
[2]Institute for Research in Electronics and Applied Physics, University of Maryland, College Park, MD 20742, USA
[3]Young Investigator Group Active Materials and Interfaces for Stable Perovskite Solar Cells, Helmholtz-Zentrum Berlin für Materialien und Energie, Kekuléstraße 5, 12489 Berlin, Germany
[4]Department of Chemical, Materials and Production Engineering, University of Naples Federico II, Piazzale Tecchio 80, 80125 Fuorigrotta, Naples, Italy
[5]Department of Chemical and Biomolecular Engineering, University of Maryland, College Park, MD 20742, USA
[6]Department of Materials Science and Engineering, University of California, Davis, CA 95616, USA

Corresponding author: mleite@ucdavis.edu

**Summary:**
Metal halide perovskite (MHP) optoelectronics may become a viable alternative to standard Si-based technologies, but the current lack of long-term stability precludes their commercial adoption. Exposure to standard operational stressors (light, temperature, bias, oxygen, and water) often instigate optical and electronic dynamics, calling for a systematic investigation into MHP photophysical processes and the development of quantitative models for their prediction. We resolve the moisture-driven light emission dynamics for both methylammonium lead tribromide and triiodide thin films as a function of relative humidity (rH). With the humidity and photoluminescence time series, we train recurrent neural networks and establish their ability to quantitatively predict the path of future light emission with <11% error over 12 hours. Together, our *in situ* rH-PL measurements and machine learning forecasting models provide a framework for the rational design of future stable perovskite devices and, thus, a faster transition towards commercial applications.

**Introduction**

Long-term stability remains the largest challenge facing the adoption of metal halide perovskite (MHP) optoelectronic devices, requiring an understanding of their fundamental material properties and the implementation of machine learning techniques that can predict their long-term performance.[1,2] Exceptional performance is no longer challenging, with MHP solar cells exceeding 25% power conversion efficiency[3] and light-emitting diodes (LED) beyond 21% external quantum efficiency (EQE)[4]. Still, commercial adoption will remain limited if the operational lifetime of perovskite photovoltaics cannot be raised to ≥25 years.[5] Unlike ubiquitous Si-based technologies, perovskites display dynamic optical and electronic responses when exposed to standard operating parameters,[1,6] including light,[7,8] temperature,[9] bias,[10] oxygen,[11-13] and water.[14-16] Among these properties, photoluminescence (PL) is critical in elucidating the proportions of radiative and nonradiative charge carrier recombination within MHP materials. For example, incident light has been proven to rapidly increase PL intensity, indicating a reduction in trap states.[17,18] Temperature-dependent time-resolved PL (tr-PL) has shown that elevated temperatures increase the rate of charge trapping and reduce the diffusion length.[19] The application of an electric field leads to iodide migration creating new sites for nonradiative recombination in polycrystalline $CH_3NH_3PbI_3$ ($MAPbI_3$).[10] The adsorption of oxygen by $MAPbI_3$ has been shown to increase the PL quantum yield (PLQY), with the kinetics depending on the grain size and repetition rate of the light source.[12] Finally, ambient moisture increases the luminescence of both formamidinium- and methylammonium-based MHP absorbers due to the passivation of surface trap states.[14,20-22] The reversibility of the dynamics depends on the specific combination and magnitude of these environmental stressors. Given that the open-circuit voltage ($V_{OC}$) is directly influenced by the amount of nonradiative recombination and still limits most MHP photovoltaics,[23-26] it is critical to understand the impact of the aforementioned environmental factors on radiative light emission.

Exposure to water can induce both chemical and structural changes in MHP materials that alter their optical and electronic behavior, requiring *in situ* measurement techniques with precise environmental control.[20,27] The ambient moisture concentration, in particular, alters the structure of the MHP lattice due to hydration reactions,[28,29] leading to changes in the extinction coefficient.[30,31] Further, water exposure may increase or decrease radiative recombination[20,21] but ultimately leads to exponentially decaying power conversion efficiency even in oxygen-free environments.[6] Density functional theory calculations have been used to explain the sensitivity to moisture, identifying preferential absorption at MA-terminated surfaces[32] and a reduction in the formation energy of hydrated phases for specific orientations of the organic cation within (001) surfaces.[33] Though it is clear for standard polycrystalline MHP films that the local stoichiometry[34] and surface termination[32] can alter water uptake, the extent of water adsorption remains an open question with conflicting reports.[35,36] More importantly, the impact of hydration and dehydration on radiative recombination, especially for different excitation photon energies, remains unknown.

Fundamental insight into MHP photophysics paired with the development of time series forecasting models will enable more rapid integration of renewable energy technologies into electric grids worldwide. Recent discussions of standard assessment protocols have explicitly called for the development of open MHP data repositories in order to build machine learning (ML) models that can identify meaningful stressors and predict degradation conditions.[2] ML techniques have already proven fruitful in other energy-related disciplines, able to predict battery cycle lifetimes based on their discharge voltage curves[37] or electrochemical impedance spectra,[38] as well as quantify Si-PV power output based on photographs of sky conditions.[39] Further, they have been applied to both guide the synthesis strategy for quantum dots[40] and compute the decay rate distributions of their emission,[41] among other applications. Concerning MHP materials, ML has been modestly used thus far, despite its tremendous potential. Examples of ML applied to perovskite materials/devices include classifying the effectiveness of post-deposition amine treatment[42] and predicting the decay in charge carrier diffusion length for different environmental stressors.[43] Additionally, MHP material synthesis has been accelerated through the application of ML techniques to determine the dimensionality and space group of novel perovskite compositions,[44,45] optimize the hole mobility of spiro-OMeTAD with a self-driving laboratory,[46] and discover new perovskite single crystals.[47] Given the ongoing search for stability, there is a strong need for accurate ML approaches that can predict the change in physical behavior (including its pathway) of any perovskite properties under environmental stress.

Here, we determine the impact of ambient moisture on light emission in both MAPbBr$_3$ and MAPbI$_3$ perovskites and build recurrent neural networks machine learning (ML) models to quantitatively predict their unique dynamics for 12+ hours. Using *in situ* PL, we measure the impact of relative humidity (rH) in both MHP compositions, finding that the I-containing material more easily adsorbs water, yet both halides exhibit moisture-driven PL enhancement. Multiple humidity cycles up to 70% rH reveal that the increase in PL intensity for pure-Br MHP is delayed (with respect to the increase in rH), establishing the existence of chemical reactions highly dependent on water partial pressure. Conversely, MAPbI$_3$ exhibits an abrupt spike in PL at ≈35% rH due to physisorption followed by quenching driven by the hydration phase transformation. The rH-cycling experiments establish negligible PL hysteresis for MAPbBr$_3$, with the ≈15× larger values presented by MAPbI$_3$ underscoring the ease at which it adsorbs water. We then use our PL data to train Echo State Networks for each composition in order to predict future light emission with an average normalized root-mean-squared error (NRMSE) <11% for a 12-hour prediction window solely based on rH input. Finally, we use a Long Short-Term Memory network to demonstrate the ability of recurrent networks to forecast the convolution of cyclical and long-term PL trends, such as moisture-induced degradation due to humidity cycling. Our work provides an extensive overview of the relationship between humidity and charge carrier recombination in Br- and I-containing perovskites, outlining both standard experimental series that should be applied to all emerging MHP compositions. Moreover, we demonstrate how ML models for time series prediction can guide the search for material stability. We anticipate our paradigm to be expanded to assess the influence of the other stress factors, including their combined effects on material properties and, ultimately, photovoltaic device performance.

Our paradigm to apply ML to predict the dynamic luminescent behavior of halide perovskites is depicted in Figure 1. First, we measure environmentally dependent PL, where the ambient humidity is the sole operational stressor and acts to modulate the perovskite light emission. As illustrated in Figure 1a and b, these experiments generate data input to build recurrent neural network (RNN) models to be used for forecasting. After data acquisition, both rH and PL time series are partitioned into the train and test datasets used to respectively develop and evaluate the RNN models, such Echo State Networks and Long Short-Term Memory, as will be discussed later. After training, the networks are utilized to predict PL output for multiple hours solely based on the registered humidity level (Figure 1c). While we chose to focus on the role of humidity in this Article, the use of RNN generalizes to multiple material properties (light absorption, crystal structure, etc.) and environmental stressors (oxygen, temperature, bias, etc.). Below, we describe the experiments used to generate the rH-PL time series used to build the predictive models.

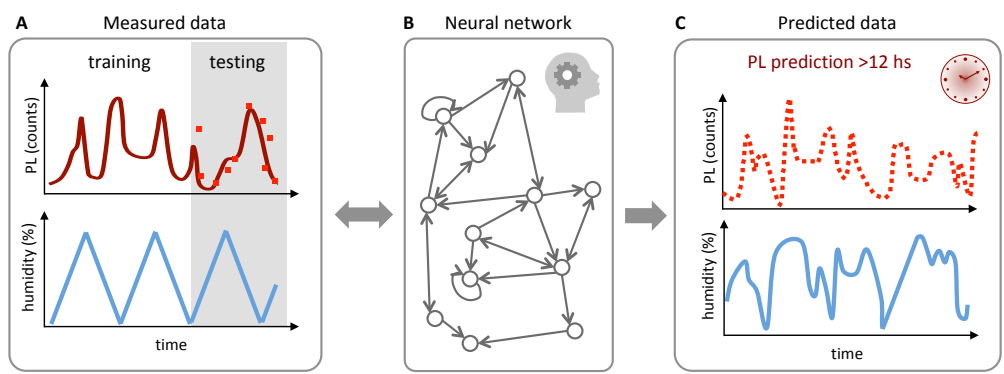

**Figure 1: Predicting perovskites' behavior with machine learning.** (a) Environmentally-dependent photoluminescence (PL) measurements are used to generate experimental input to build recurrent neural network (RNN) models used for forecasting. In this particular case, the ambient relative humidity (rH) is the sole operational stressor and acts to modulate the perovskite light emission. After acquisition, the rH and PL time series are split into train and test datasets and used to (b) train the RNN models, such Echo State Networks and Long Short-Term Memory. (c) After training, the networks are utilized to predict PL output for multiple hours solely based on the registered humidity level.

**Results**

### Moisture-driven photo-emission dynamics

First, we resolve the MAPbI$_3$ and MAPbBr$_3$ light emission dynamics at constant relative humidity levels, establishing the relationship between water adsorption and magnitude of radiative recombination (see Figure 2). All sample fabrication and experimental preparation took place inside a nitrogen-filled glove box to exclude uncontrolled material changes/degradation. We use micro-PL measurements based on a confocal optical microscope to capture the light emission dynamics, controlling the sample environment with a programmed mass flow controller (MFC) that adjusts the ratio of dry and moist N$_2$ flowing through the microscopy chamber (Figure S1). Films equivalent to those used for the PL experiments were incorporated into solar cells, demonstrating the high quality of the materials (see Experimental Methods for both thin-film and device fabrication details). The champion MAPbI$_3$ and MAPbBr$_3$ devices present 15.4% and 8.2% reverse-scan direction power conversion efficiencies under AM1.5G 1 sun illumination (Figure S2), respectively.

For all relative humidity (rH) conditions, the Br-containing sample exhibits PL intensity decay during the initial 15 min period under inert conditions (see Figure 2a). The 19% and 41% rH data (Figure 2c and e) exhibit no noticeable features while soaking in the moist environment. Higher water concentrations are required to alter the sample's luminescence, as indicated by the 59% rH measurement shown in Figure 2g. Here, the PL enhancement due to physisorption lags approximately 14 min behind the onset of the maximum rH level. The spike in luminescence intensity is followed by a gradual increase over the next ≈3.5 h as surface trap states are passivated, possibly due to recrystallization[48] or chemisorption (as reported in MAPbI$_3$[14]). Finally, when the sample returns to inert conditions (rH < 50%), the PL dramatically increases.

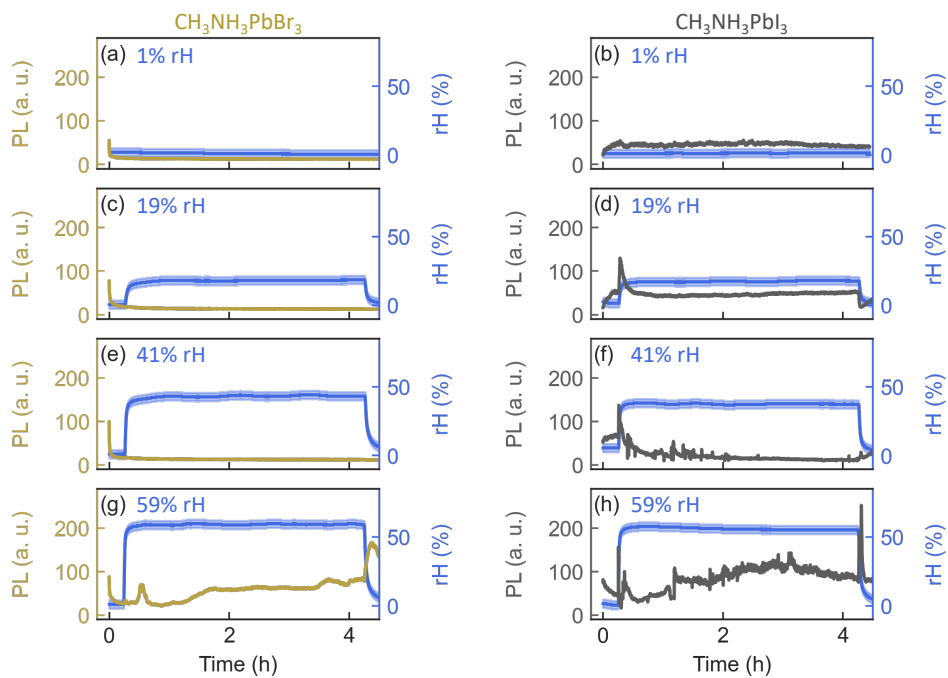

**Figure 2: Light-emission dynamics of halide perovskites at constant relative humidity.** Photoluminescence as a function of time for MAPbBr$_3$ (left - gold) and MAPbI$_3$ (right - grey) at stabilized (a, b) 1%, (c, d) 19%, (e, f) 41%, (g, h) 59% rH with an accuracy of ±4% rH. The rH sensor (blue) has a resolution of ±3% rH (shaded blue area). The values indicated on the plots are the average of the rH values for both compositions. Excitation wavelength: 532 nm.

Starved of water, the chemical reactions at the sample surface stop, while the small amount of water temporarily remaining at the surface leads to increased radiative recombination.[21] Once inert conditions are restored, the chemical potential gradient established between the film and its atmosphere dislodges the water at the MAPbBr$_3$ surface. This final dehydration of the Br-containing perovskite is marked by the peak and subsequent decay in PL intensity, as shown after 4 hours of measurement in Figure 2g.

MAPbI$_3$ exhibits much greater sensitivity to water, revealed by modulated PL dynamics even at low humidity levels (19% rH – Figure 2d). The first 15 minutes of measurement reveal a modest increase in light emission (Figure 2b, d, f), as widely reported in the literature.[17,18] Then, the rise in PL onsets as soon as the rH sensor registers an increase in moisture concentration. For all humid conditions, the spike in PL is followed by an immediate exponential decay. The reduction in the duration of the PL intensity surge with increasing moisture provides direct evidence of the narrow processing window for moisture-based passivation treatments of MAPbI$_3$. The measurement at 41% rH (Figure 2f) displays a similar exponential decay but stabilizes 70% lower than the 19% rH series (Figure 2d). Once the water partial pressure is sufficient, the entire perovskite surface may be covered with water molecules, accelerating radiative recombination.[49] Only in the case of the 59% rH condition (Figure 2h) are surface chemical reactions able to begin passivating non-radiative trap states and increase luminescence. Soaking the sample at 59% rH causes a slower light emission enhancement, but eventually raises the PL beyond their initial values due to passivating chemical reactions (Figure 2h, hours 1-3). Together, the results for 19% (Figure 2d), 41% (Figure 2f), and 59% rH (Figure 2h) indicate that the water vapor pressure directly impacts how much water is absorbed by the perovskite film and, thus, the speed of any chemical reactions and subsequent phase transformations.[30] All humidity levels show abrupt changes when the moisture is removed from the sample environment. The 19% and 41% rH conditions (Figure 2d and f, respectively) show nonexistent or negligible peaks, while the 59% rH data (Figure 2h) reveals an abrupt spike representing the highest registered PL value across the entire set of constant-rH measurements.

Both I- and Br-based perovskites react similarly to water, beginning with the physisorption of H$_2$O molecules passivating surface trap states and increasing overall radiative recombination.[14,21] However, once the water partial pressure is sufficient, the entire surface of the perovskite becomes covered with water molecules that act as *n*-type dopants and diminish carrier lifetime.[49-51] Our measurements (Figure 2c, e, and g) demonstrate that trap states in the Br-MHP material require >45% rH for measurable passivation before displaying the anticipated decay in luminescence. Prior experimental investigations into MAPbBr$_3$ single crystals have also shown that passivating these surface trap states dramatically increases the PL lifetime,[21] implying that the PL intensity enhancement we register in Figure 2g is due to both increasing lifetime and overall yield. As our results support, water-driven changes of MAPbBr$_3$ photoemission require even higher vapor concentrations for notable adsorption, given the smaller lattice constants (b-axis bond length: Pb-I = 3.249 Å, Pb-Br = 3.031 Å[7,52]) and correspondingly higher energies for interacting with water.[53] By contrast, MAPbI$_3$ is passivated more readily, with 19% rH (Figure 2d) revealing a brief passivation-related increase in radiative recombination before returning to the initial level of emission. Compared to 41% (Figure 2f) and 59% rH (Figure 2h), there is insufficient water to coat the entire surface and decrease charge carrier recombination. Finally, our PL measurements indicate that chemical reactions at the film surface gradually increase overall light emission for both compositions (Figure 2g and h). In the case of MAPbI$_3$, this observation is explained by moisture-induced degradation of the first few nanometers of the film into PbI$_2$.[14]

Next, we establish the reversibility and repeatability of the light emission dynamics across multiple humidity cycles (0-70% rH), with both compositions displaying almost complete luminescence recovery across three cycles (Figure 3). We first consider the Br-containing perovskite (Figure 3a) which displays the characteristic lag in the initial short-lived 15× PL rise, followed by a modest 3.4× increase at 0.7 h. Upon further cycling, the initial signal increase at times 1.8 and 3.2 h barely register, while the slower process at 2.2 and 3.4 h becomes more pronounced. We suggest that the initial PL increase corresponds to physisorption, while the latter is related to chemisorption. The PL intensity enhancement is limited during the third humidity cycle likely due to a combination of (i) the smaller lattice constant for MAPbBr$_3$ limiting water's penetration depth and (ii) prior surface chemical reactions increasing the density of shallow trap states.[50] The changes in PL for MAPbI$_3$ (Figure 3e) occur more abruptly. The introduction of moisture leads to a PL peak around ≈35% rH, after which the characteristic decay is observed. We attribute the decrease in radiative recombination to (i) the full coverage of the MAPbI$_3$ surface with water molecules creating new trap states (as discussed above) and (ii) the eventual formation of a monohydrate phase with band gap of 3.1 eV that cannot be excited with our 2.3 eV (532 nm) laser source used in our experiments.[30] The volume of the film converted to MAPbI$_3$•H$_2$O, or even MAPbI$_3$•2H$_2$O depends on the ambient water vapor pressure and duration of exposure, with moisture adsorption quickly plateauing.[35] Given the transparency of the hydrated MAPbI$_3$ layer, the charge carrier generation shifts deeper into the film according to the proportions of the different phases.[50] The changes in crystal structure also promote new nonradiative recombination sites within the interaction volume ($\frac{1}{\alpha} \approx 165\,nm$)[54] of the laser. However, these changes are reversible, as evidenced by the increased PL signal after the film dries out (Figure 3e). Additional PL time series acquired over three humidity cycles for both MAPbBr$_3$ and MAPbI$_3$ are shown in Figure S3 and S4, respectively.

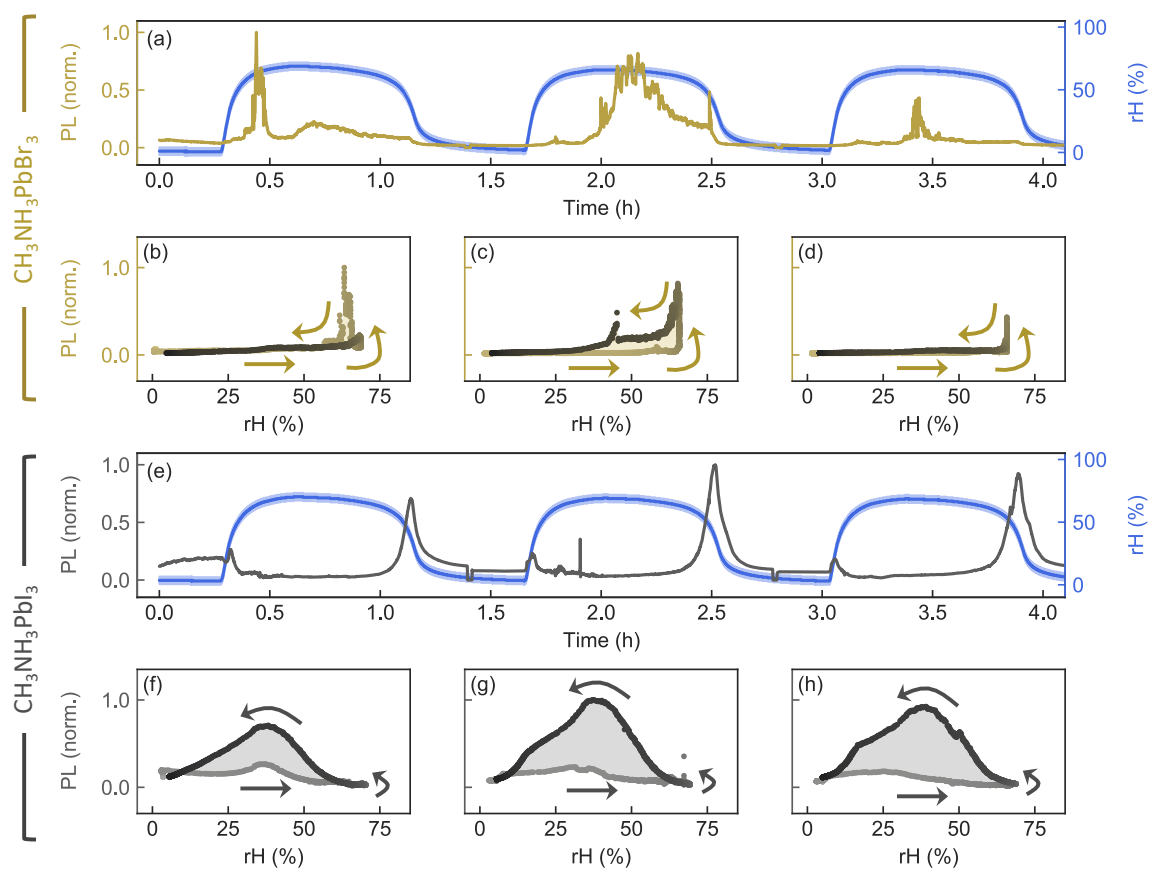

**Figure 3: Humidity cycling drives reversible PL behavior in MAPbBr$_3$ and MAPbI$_3$.** (a, e) Photoluminescence as a function of time and the extracted first, second, and third rH-PL hysteresis loops for (b-d) MAPbBr$_3$ and (f-h) MAPbI$_3$, respectively. In (a) and (e), the dark blue line refers to the measured rH value while the shaded blue area represents the ±3% rH resolution. The adjacent arrows (a-d, f-h) and color gradient both indicate the time evolution of the rH-PL curves. Excitation wavelength: 532 nm.

The correspondence between relative humidity and light emission (Figure 3b-d and f-h) can be extracted from the time-series data, revealing the extent of hysteresis and the rH-PL loop area for each sample. Here, we define the hysteresis for each rH-PL cycle as the difference between the initial and final measurements, enabling us to quantify the reversibility of the reactions driving the PL dynamics. Both compositions show minimal hysteresis across all cycles, unimpacted by the cumulative duration of moisture exposure. Unsurprisingly, MAPbI$_3$, with a more open crystal structure, exhibits ≈15× larger average values compared to MAPbBr$_3$ (Figure S5). The area enclosed within the rH-PL loop indicates the extent and rate of the reversible chemical reactions (see below), *i.e.* how strongly water has impacted the material. For all three cycles measured on the pure-Br perovskite, the rH-PL loop area only grows above 40% rH. The first and third cycles enclose a negligible area. By contrast, 70% rH is sufficient to initiate a progression of hydration reactions for MAPbI$_3$:

$$CH_3NH_3PbI_3 + H_2O \leftrightarrow CH_3NH_3PbI_3 \cdot H_2O \qquad (1)$$

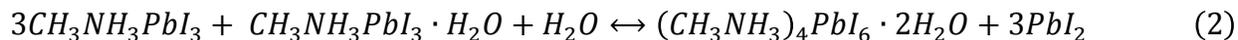

$$3CH_3NH_3PbI_3 + CH_3NH_3PbI_3 \cdot H_2O + H_2O \leftrightarrow (CH_3NH_3)_4PbI_6 \cdot 2H_2O + 3PbI_2 \quad (2)$$

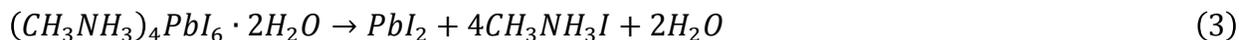

$$(CH_3NH_3)_4PbI_6 \cdot 2H_2O \rightarrow PbI_2 + 4CH_3NH_3I + 2H_2O \quad (3)$$

where water molecules drive the formation of monohydrate and then dihydrate compounds.[29,30] The dihydrate phase can then irreversibly convert into PbI$_2$ and MAI, leaving the water molecules behind to continue the hydration process. The larger lattice parameters for MAPbI$_3$ render it more easily affected by water, leading to significant PL hysteresis (Figure 3f-h). The rH-dependent hysteresis demonstrates that the rate of hydration is much faster than that of dehydration. The moisture-driven changes to the MAPbI$_3$ film were investigated with both optical and scanning electron microscopy (SEM) (Figure S6a-r), revealing that water corrodes the surface heterogeneously. The optical micrographs were acquired after each of six consecutive, identical humidity cycles. As the duration of exposure increases, so does the area percentage of the film covered in black spots. To verify, we acquire SEM with energy-dispersive X-ray spectroscopy (EDS) images, and the Si, Pb, and I signal intensity (Figure S6p-r and Figure S7a-d, respectively) reveal an anticorrelation between Si and Pb/I concentrations.

**Predicting optical properties through Machine Learning**

We develop a series of recurrent neural network ML models to predict PL dynamics in halide perovskites. We use Echo State Networks (ESN), a type of recurrent neural network, to forecast the PL signal across multiple cycles, solely based on the recorded humidity during that period (see Supplementary Information for an extended overview of the training details). ESN are an ideal choice for time series prediction because they contain dynamic memory, allowing them to replicate the behavior of nonlinear systems.[55] Further, they are easily implemented and have greater computational efficiency compared to other recurrent neural networks, such as gated recurrent units or long short-term memory (LSTM). ESN have proven capable of forecasting wind speeds for turbine generation,[56] the magnitude of solar irradiance governing PV power output,[57] and the remaining lifetime of both fuel cells[58] and batteries.[59] ESN have also been applied successfully in other fields, including time series predictions of chaotic systems, such as nonlinear time delay (Mackey-Glass)[55] and diffusive instability (Kuramoto-Sivashinsky).[60]

The ESN is trained using rH-PL data obtained over three rH cycles (Figures 2, S3, and S4). To increase model accuracy, the dataset is enlarged through a straightforward augmentation routine (complete details in the Supplementary Information), wherein additional time series are simulated by interpolating between the various ground truth experimental data (Figure S8).[61,62] The full time series consisting of both experimental and simulated data is then generated by taking random combinations of all the data. We assume that no material degradation occurs, consistent with the occurrence of higher moisture-induced PL values beyond the first rH cycle (Figure 3e). We note that the use of random combinations will not be valid for time series prediction of long-term MHP degradation over months and years, where the temporal order plays a central role (addressed further below). The overall randomly generated time series is split 80%-20% into train and test datasets (see Figure S9 for network structure). Figure 4a and d shows the test set performance of the ESN models for $MAPbBr_3$ and $MAPbI_3$, respectively. We quantify model performance with the normalized root-mean-squared error (NRMSE) metric:

$$NRMSE = 100 \times \frac{\sqrt{\sum_{i=1}^{n}(y_{observed}-y_{predicted})^2/n}}{\max(y_{observed})-\min(y_{observed})} \qquad (4)$$

where $y_{observed}$ is the PL data from the experiments, $y_{predicted}$ is the predicted PL from the trained network, and *n* is the number of samples in the dataset. Importantly, NRMSE is scale-invariant, allowing comparison between all models. The NRMSE for the test set is only 10.5% and 7.4% for pure-Br and pure-I, respectively. These low error levels are measured over 12.2 h, for MAPbBr$_3$ and MAPbI$_3$, respectively, only relying on humidity input and without periodically retraining the model. The ESN performance optimization relies on a grid search approach (Figure 4b and e), varying the noise and spectral radius of the network. The low variance across the heatmaps indicates that the network predictions are stable across a wide hyperparameter space, as desired. The overall PL trends from both models reveal a strong qualitative agreement. The MAPbBr$_3$ displays worse predictive performance, which is understandable given the sporadic behavior of the experimental data. The trajectory of the MAPbI$_3$ model is rarely, if ever, inversely correlated with the experimental data. The primary challenge for the model applied to MAPbI$_3$ lies in quantitatively predicting the full extent of the moisture-driven PL increases. We attribute this behavior to two causes: (i) substantial sample-to-sample variation in the intensity of PL enhancement and (ii) increases in PL emission for these humidity conditions are short-lived, leaving fewer data points in their vicinity. Further investigation is needed to understand the sample-to-sample variations, an issue with implications that extend beyond the rH-PL measurements detailed in this study. The spikes in light emission might be more accurately computed by (i) rebalancing the distribution of PL values with a combination of synthetic undersampling and oversampling[63,64] or (ii) using an error metric with heavier penalties for under predicting (*e.g.* root-mean-squared logarithmic error).

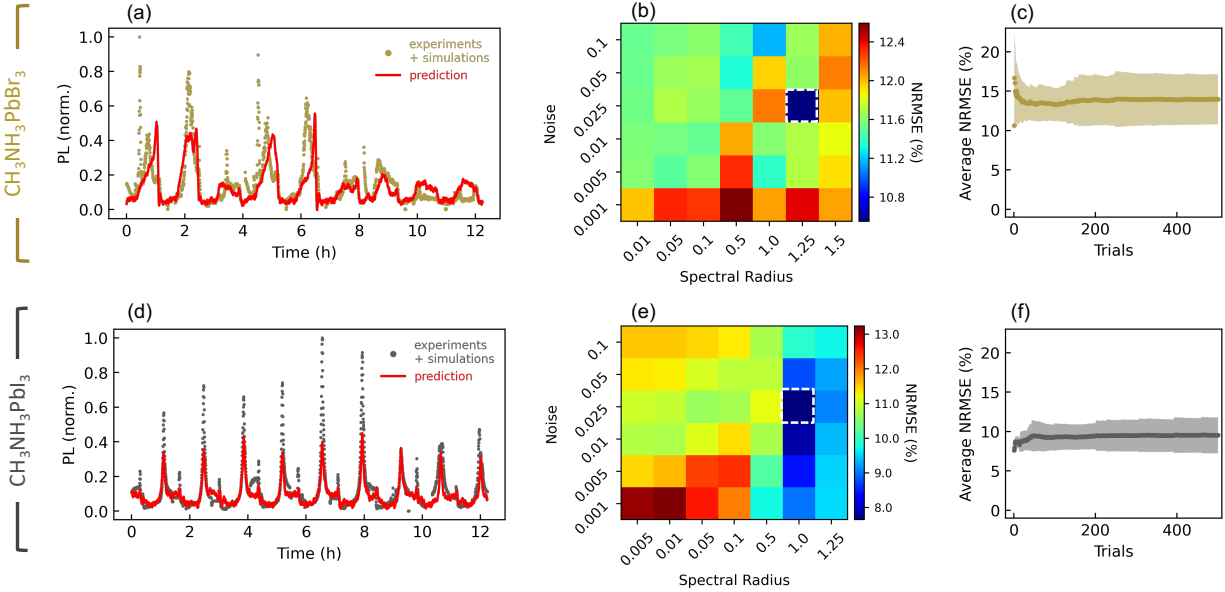

**Figure 4. Echo State Networks (ESN) enable accurate PL forecasting solely based on the ambient humidity.** (a, d) Augmented rH-PL experimental data ($MAPbBr_3$ = gold markers, $MAPbI_3$ = grey markers) and ESN predictions (red line) as a function of time. An 80%-20% train-test split is used for both compositions. Heatmaps of NRMSE for (b) $MAPbBr_3$ and (e) $MAPbI_3$ outline the grid-search approach to varying hyperparameters (noise and spectral radius) to optimize ESN predictions on the test set. The white dashed boxes indicate the hyperparameters used for the predictions in (a) and (d). (c, f) Average NRMSE as a function of the number of trials with the standard deviation indicated by the shaded region. For both $MAPbBr_3$ and $MAPbI_3$, the fluctuations in average NRMSE vanish beyond 100 trials.

Next, we demonstrate that the low error rate of the network is independent of the order of the PL cycles by testing performance across a wide range of trials (Figure 4c and f). Here, we define a trial as one randomly generated sequence of experimental and simulated rH-PL data. While each trial is produced from the same set of experimental and simulated rH-PL data, the order is set by choosing unique random seeds. Figure 4c and f displays the average NRMSE for a given number of trials and the corresponding standard deviation is indicated by the shaded region. For both compositions, the average NRMSE is below 14%. However, the standard deviation after 500 trials is noticeably different at 3.2% and 2.3% for $MAPbBr_3$ and $MAPbI_3$, respectively. Thus, for the I-containing films, 420 out of 500 trials (84%) present NRMSE

<12%. We attribute the more accurate average performance for MAPbI$_3$ to the consistency of the general trends in PL dynamics across samples (see Figure S4 for additional *in situ* PL measurements).

We further use recurrent neural networks to predict the convolution of "seasonal" dynamics along with long-term PL trends of MAPbI$_3$ degradation, as shown in Figure 5. To approximate water-induced degradation, the experimental and simulated PL arrays are stitched together in descending order of maximum PL values. As previously reported, over the long-term (multiple weeks), water degrades the power conversion efficiency of MAPbI$_3$ solar cells through the formation of hydrated species that may decompose into PbI$_2$.[28-30] Our microscopic investigations into film morphology and chemical composition (Figure S6) provide direct evidence of grain corrosion that worsened over six humidity cycles (≈8.2 h). The PL dynamics remained pronounced after six cycles, yet, based on prior research,[28,31,65] we expect an overall diminishing trend to become dominant over 72 cycles (≈94 h). To predict the experimental time series shown in Figure 5a, we implement a long short-term memory (LSTM) network, another type of recurrent neural network. The network is trained to predict the PL value three humidity cycles in the future (≈4.1 h) based on the PL and rH data from the previous three cycles. The ≈4.1-hour lag between cycles accounts for the gap between the train and test set predictions. After training for 20 cycles, the LSTM model achieved an NRMSE of 4.6% and 1.8% for train and test partitions, respectively (Figure 5b). The model predictions displayed in Figure 5a indicate that the LSTM can quantitatively forecast the path of future PL values. Rather than rely on the random initialization of network weights, as with ESN, we tune the weights appropriate to the rH-PL data over many epochs of training (Figure 5b). We chose LSTM over ESN for the simulated degradation task, despite the increased computational expense, in order to train all of the network weights for the prediction. After training, the LSTM now retains prior data points in memory for the appropriate amount of time for our scenario. A zoom-in view of the test set predictions is shown in Figure 5c, revealing good agreement with the (i) abrupt spikes occurring at ≈35% rH, (ii) quenching at 70% rH, and (iii) the overall trend of decreasing intensity. To demonstrate the difference in performance, we apply the same ESN model previously implemented for MAPbI$_3$ (Figure 4c) to the degradation scenario (Figure 5d).

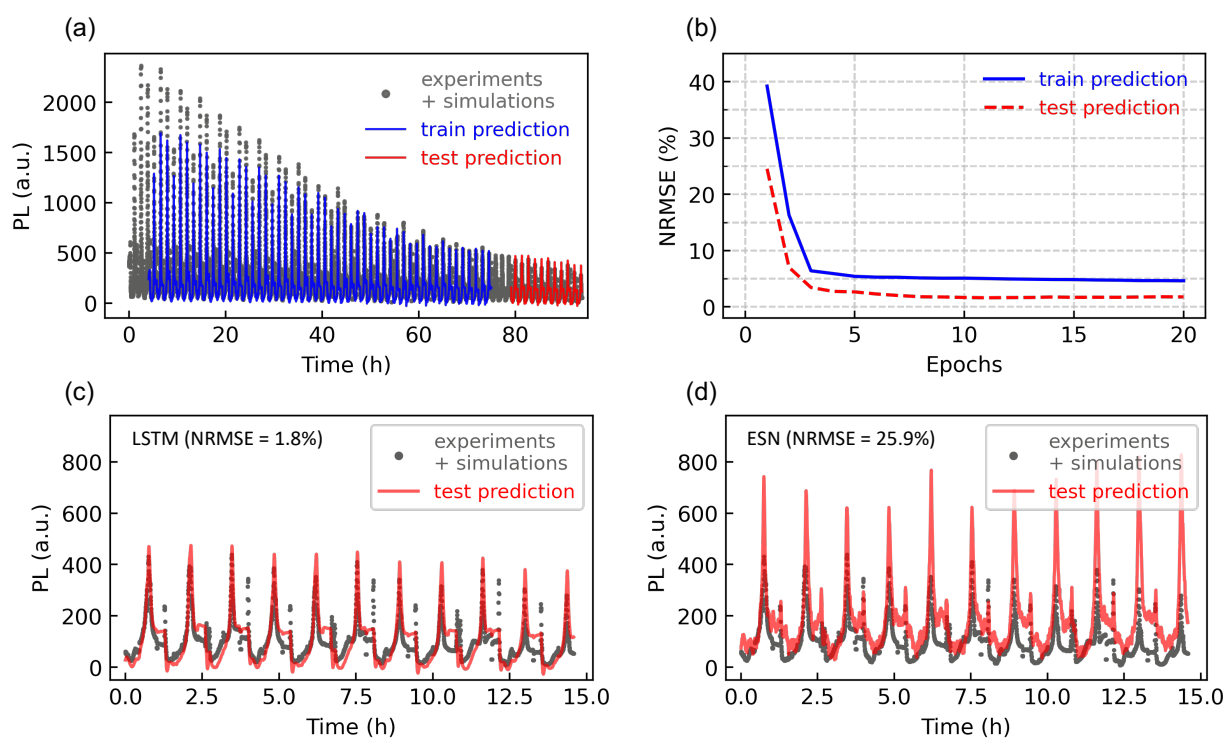

**Figure 5. Predicting degradation in MAPbI$_3$ with machine learning.** (a) The MAPbI$_3$ augmented experimental data (grey) simulating degradation compared against train (blue) and test (red) predictions from an LSTM network. (b) The NRMSE loss function for the train (blue solid line) and test (red dashed line) partitions over the course of 20 epochs shows <5% error for both. The LSTM network is trained to make rolling predictions off the future PL value in three humidity cycles (≈4.1 h) based on the PL and humidity for the past three cycles. The performance of (c) LSTM and ESN (d) models on the test portion of the simulated degradation task. Unlike, the ESN, which continues to overpredict the magnitude of the moisture-driven PL rises, the LSTM has learned the overall decreasing trend. The lower NRMSE (1.8%) obtained by the LSTM demonstrates its more adaptable memory.

The approaches outlined by our work are easily adapted to additional arrangements of environmental stressors, enabling forecasts for unexplored combinations of oxygen, water, light, temperature, and bias.[1] By building large shared data repositories that include information on fabrication methodology and *in situ* chemical information, ML techniques will enable the development of accurate metrics that indicate degradation hours, days, or even weeks in advance. Clarifying which fundamental properties exert the most influence over long-term material properties will lead to more targeted compositional engineering which, in turn, will accelerate the development of stable perovskite for optoelectronics.

**Discussion**

We resolved the moisture-induced luminescence dynamics in the archetypal $MAPbBr_3$ and $MAPbI_3$ compositions and demonstrated the ability of machine-learning models based on recurrent neural networks to forecast the time evolution of light emission hours into the future. Through constant humidity experiments, we established the vapor pressure thresholds for both transient PL escalation and diminution for each composition. We identified reversible and repeatable luminescence enhancement $\geq 6\times$ for both halide species by *in situ* rH-PL cycling experiments. Our measurements reveal the impact of the halide species in dictating water ingress, with $MAPbBr_3$ showing insignificant rH-PL hysteresis, $\approx 15\times$ less than $MAPbI_3$. For the pure-I MHP, the extent of luminescence increase grows over subsequent cycles despite ongoing degradation at the sample surface. Importantly, our time-dependent PL allowed us to elucidate the existence of both physisorption and chemisorption events for $MAPbI_3$ perovskites. We then implemented recurrent neural networks to compile a predictive model for the rH-driven dynamics, using ESN to quantify the exact path of the PL dynamics yielding an error rate <11% over 12 hours of prediction. Finally, we demonstrated an LSTM network capable of predicting the long-term light emission trend of a MAPbI$_3$ sample degrading over 72 humidity cycles, reaching percentage errors of 4.6% and 1.8%, for train and test sets, respectively.

The environmentally-controlled PL measurements presented here provide a robust framework applicable to *any* perovskite composition that can be extended to include other environmental stressors, such as temperature and bias, and material properties, such as crystal structure, light absorption, charge carrier lifetime, and electrical conductivity. Specifically, *in situ* X-ray diffraction (XRD),[30,66] spectroscopic ellipsometry,[54,67] time-resolved PL,[14,22] and microwave photoconductivity[68,69] could be applied to perovskites submitted to similar rH cycles. All of these techniques are contactless, allowing assessment of the perovskite material without the need to consider the role of interfaces due to the addition of metallic contacts. Our ML models could be extended to these measurements, where the accuracy of recurrent neural networks, such as ESN and LSTM, would be tested against different data features. Similar data transformations could be applied to other environmental stressors and material properties, identifying the primary factors governing the perovskite dynamic response. For example, aging tests considering elevated temperature (>50 °C) under varying moisture concentrations or different atmospheres (O$_2$ or N$_2$), as well as distinct light treatments, could be implemented and broaden to consider their combined effects in the perovskites' response. By measuring a wide array of material properties under identical operating conditions, the subtle but relevant relationships between material structure and optoelectronic properties may be observed and used to predict device behavior and performance. Thus, we foresee the implementation of ML-based diagnostic tools, where new thin-film perovskite compositions could be rapidly and accurately screened for long-term stability based on contactless measurements.

**Experimental Procedures**

Fabrication of thin films and devices:

*Cleaning of substrates*: Before usage, pre-patterned FTO substrates (TEC 15, Yingkou company, 2.5 cm × 2.5 cm) and microscopic glass were cleaned by sonication first in 2% Mucasol solution (Schülke company) in water, and then acetone and isopropanol for 15 min, respectively. Then the substrates were cleaned for another 15 minutes using UV-ozone treatment before usage.

*Deposition of $TiO_2$ layer*: Compact $TiO_2$ layer was deposited at 450 ºC *via* spray pyrolysis where the sprayed solution is made of 0.48 ml of acetylacetone and 0.72 ml of titanium diisopropoxide bis(acetylacetonate) in 10.8 ml of ethanol for 24 substrates in an array of 4 columns by 6 rows. Mesoporous $TiO_2$ layer was deposited from a diluted $TiO_2$ paste (30 N-RD $TiO_2$ paste, Great Cell) at the concentration of 120 mg per ml of ethanol at the spin-coating speed of 4000 rpm for 10s. After that, the samples were annealed at 450 ºC for 30 min.

*Thin film deposition of perovskites*: The solutions were prepared and deposited inside nitrogen-filled glove boxes.

*$MAPbBr_3$:* 1.24 M of $MAPbBr_3$ precursor solution was prepared from 0.15 g MABr dissolved with 1042 μL 1.5 M $PbBr_2$ stock solution in DMF: DMSO mixture solvent (volume ratio of 4:1) and 38 µL mixed solvent at the molar ratio of 1:1.09. Then 60 µL was deposited on the substrate. The spin-coating step involves first at 1000 rpm with an acceleration speed of 200 rpm/s and a dwell time of 10 s, and then 4000 rpm with an acceleration speed of 2000 rpm/s and a dwell time of 30 s. 100 µL of chlorobenzene was added on top of the substrate 15 s before the end of the spin-coating program. Then the film was placed on a hot plate at 100 ºC for 15 min.

*MAPbI$_3$:* 1.2 M of MAPbI$_3$ precursor solution was prepared from 0.2 g MAI dissolved with 1041 µL 1.3 M PbI$_2$ stock solution in DMSO and 11 µL pure DMSO solvent at the molar ratio of 1:1.09. Then 60 µL was deposited on the substrate. The spin-coating step involves first at 1000 rpm with an acceleration speed of 200 rpm/s and a dwell time of 10 s, and then 6000 rpm with an acceleration speed of 2000 rpm/s and a dwell time of 30 s. 100 µL of chlorobenzene was added on top of the substrate 15 s before the end of the spin-coating program. Then the film was placed on a hot plate at 100 ºC for 40 min.

*Deposition of hole transport layer:* 50 mg of spiro-OMeTAD was dissolved in 1383.1 µL chlorobenzene. Then 19.92 µL 4-tert-Butypyridine, 12.17 µL of lithium bis(trifluoromethanesulfonyl)imide (LiTFSI) from its stock solution 520mg/ml in acetonitrile, and 20.19 µL of tris(2-(1H-pyrazol-1-yl)-4-tert-butylpyridine)cobalt(III) tri[bis(trifluoromethane)sulfonimide] (FK209) from its stock solution 300 mg/ml in acetonitrile were added into the solution. Then the doped spiro-OMeTAD solution was spin-coated on top of perovskite covered substrate at the speed of 1800 rpm with an acceleration rate of 200 rpm per second and a dwelling time of 30 seconds. Then, the samples were stored inside a dry air box with a constant dry air flow overnight.

*Deposition of Au electrode:* The samples were transferred into a thermal evaporator for the deposition of 80 nm gold at the vacuum of below $1\times10^{-6}$ bar. The device active area is 0.144 cm$^2$ determined by the shallow mask.

Device characterization:

The Wavelabs Sinus-70 LED class AAA solar simulator was used as the light source for the *J-V* scans. The light intensity of 100 mW cm$^{-2}$ was calibrated with a Silicon reference cell from Fraunhofer ISE. A Keithley 2400 SMU, controlled by a measurement control program written in LabView, was used for the *J-V* scans with a scan rate of 50 mV/s. 18 solar cells of MAPbBr$_3$ and MAPbI$_3$ individually were fabricated.

*In Situ* Photoluminescence:

Photoluminescence (PL) measurements of neat perovskite films deposited on microscopic glass were performed with a confocal microscope using a 100x, 0.75 N.A. objective in reflection mode and a photomultiplier tube (PMT) for detection. For all PL experiments, a 532 nm power-tunable diode laser was used with power densities of 3,500 mW cm$^{-2}$ and 14,000 mW cm$^{-2}$ for MAPbBr$_3$ and MAPbI$_3$, respectively. Prior to coupling the laser illumination into the microscope, we coarsely attenuated the power with absorptive neutral density filters, modulated the light at 150 Hz with an optical chopper, and cleaned up the background emission with a 40 nm wide (FWHM) bandpass filter centered at 550 nm. An additional 750 nm short pass filter was included for the measurements on MAPbI$_3$, while a 546 nm short pass filter was added for MAPbBr$_3$. A long pass filter blocks the reflected excitation source, leaving only the PL incident upon the PMT. Here, the long pass filter wavelengths, 750 nm for MAPbI$_3$ and 550 nm for MAPbBr$_3$, were selected to provide the maximum range of spectrally-dependent measurements. The PMT was operated at 650 V and its output was recorded with a lock-in amplifier.

Humidity control:

In all instances, the sample ambience was controlled through the use of a custom-built environmental chamber.[20] The samples are loaded into the sealed chamber inside a glovebox, preventing unintended exposure to water or oxygen. Once loaded into the microscope, a flow of dry N$_2$ is introduced. Prior to the chamber, the N$_2$ was bifurcated into wet and dry channels. The dry line has a fixed flow rate while the wet line passes through a mass flow controller (MFC) then a water bubbler. The MFC voltage setpoint was changed according to the user-defined program, allowing repeatable humidity cycles. A capacitive relative humidity sensor located less than 30 mm from the sample surface measured the ambient moisture.

Scanning Electron Microscopy:

The micrographs and energy-dispersive X-ray images were acquired with a Hitachi SU-70 FEG SEM operated at a 10 kV accelerating voltage with 6000× magnification.

Echo State Network predictions for rH-PL:

A 150-node echo state network (ESN)[55] was applied to the PL time series, using the simultaneously recorded rH data as the sole input. The network states were generated by passing through 80% of the PL and rH data, after which only the output layer readout weights were fit to minimize the mean squared error between the predicted and measured luminescence values. The data preprocessing routines, network structure, and governing mathematical equations are described fully in the Supplementary Information.

Long Short Term Memory predictions for simulated PL degradation:

A 20-node long short-term memory (LSTM) network was trained for 20 epochs using an 80%-20% train-test split. The MSE loss function and the Adam optimizer were used. The entire set of simultaneously recorded rH and PL values (Figure S4) over three cycles (4.1 h) was used to generate the augmented dataset as input data for the LSTM. The LSTM was trained to make rolling predictions of the PL value three cycles in advance based on this input data. The augmentation routine and full network architecture is described in the Supplementary Information.

**Acknowledgements**

The authors thank Prof. B. Neves for help with the environmental control setup, and N. Ballew and T. Weimar for machining assistance. MSL thanks the financial support from the National Science Foundation (EPMD award #1610833) and UC Davis. J.M.H. acknowledges the funding support from MEII's 2018–2019 Harry K. Wells Graduate Fellowship, UMD's 2019 Graduate Summer Research Fellowship, and UMD's Fall 2019 Ann G. Wylie Dissertation Fellowship.

**Author Contributions**

J.M.H. and M.S.L. conceived the experiments. Q.W. fabricated the samples. J.M.H. performed the measurements and modeling. J.M.H., E.L., R.L., and M.S.L. interpreted the results. All authors discussed the results, reviewed and edited the manuscript. M.S.L. supervised the work.

**Declaration of Interests**

There are no conflicts to declare.